\documentclass[aps,prb,floatfix,showpacs,twocolumn,superscriptaddress]{revtex4}

\usepackage[utf8]{inputenc}
\usepackage[english]{babel}
\usepackage{graphicx}
\usepackage{dcolumn}
\usepackage{amsmath}

\newcommand{\ie}{i.e.}
\newcommand{\etal}{{\it et al.}}
\newcommand{\alphaETI}{$\alpha$-(BEDT-TTF)$_2$\-I$_3$}
\newcommand{\cm}{cm$^{-1}$}

\begin{document}

\title{Semimetallic and charge-ordered $\alpha$-(BEDT-TTF)$_2$I$_3$: on the role of disorder in dc transport and dielectric properties}

\author{Tomislav Ivek}
\author{Matija \v{C}ulo}
\thanks{These two authors contributed equally.}
\email{mculo@ifs.hr}
\affiliation{Institut za fiziku, P.\ O.\ Box 304, HR-10001 Zagreb, Croatia}
\author{Marko Kuve\v{z}di\'{c}}
\affiliation{Institut za fiziku, P.\ O.\ Box 304, HR-10001 Zagreb, Croatia}
\affiliation{Department of Physics, Faculty of Science, University of Zagreb, Bijeni\v{c}ka cesta 32, 10000 Zagreb, Croatia}
\author{Eduard Tuti\v{s}}
\affiliation{Institut za fiziku, P.\ O.\ Box 304, HR-10001 Zagreb, Croatia}
\author{Mario Basleti\'{c}}
\author{Branimir Mihaljevi\'{c}}
\author{Emil Tafra}
\affiliation{Department of Physics, Faculty of Science, University of Zagreb, Bijeni\v{c}ka cesta 32, 10000 Zagreb, Croatia}
\author{Silvia Tomi\'{c}}
\affiliation{Institut za fiziku, P.\ O.\ Box 304, HR-10001 Zagreb, Croatia}
\author{Anja L\"{o}hle}
\author{Martin Dressel}
\author{Dieter Schweitzer}
\affiliation{Physikalisches Institut, Universit\"{a}t Stuttgart, Pfaffenwaldring 57, D-70550 Stuttgart, Germany}
\author{Bojana Korin-Hamzi\'{c}}
\affiliation{Institut za fiziku, P.\ O.\ Box 304, HR-10001 Zagreb, Croatia}

\date{\today}

\begin{abstract}
\alphaETI{} is a prominent example of charge ordering among organic conductors. In this work we explore the details of transport within the charge-ordered as well as semimetallic phase at ambient pressure.  In the high-temperature semimetallic phase, the mobilities and concentrations of both electrons and holes conspire in such a way to create an almost temperature-independent conductivity as well as a low Hall effect. We explain these phenomena as a consequence of a predominantly inter-pocket scattering which equalizes mobilities of the two types of charge carriers.  At low temperatures, within the insulating charge-ordered phase two channels of conduction can be discerned: a temperature-dependent activation which follows the mean-field behavior, and a nearest-neighbor hopping contribution. Together with negative magnetoresistance, the latter relies on the presence of disorder. The charge-ordered phase also features a prominent dielectric peak which bears a similarity to relaxor ferroelectrics. Its dispersion is determined by free-electron screening and pushed by disorder well below the transition temperature. The source of this disorder can be found in the anion layers which randomly perturb BEDT-TTF molecules through hydrogen bonds.

\end{abstract}


%
%
%

\maketitle


\section{Introduction}

The quest for exotic electronic orderings is driven by our ability to produce said states as clean as possible, in a tunable manner and preferably cheap. Organic conductors with reduced dimensionality offer all of these desirable properties and are rightly in the very focus of solid-state physicists searching for new phenomena. The origin of their extremely rich phase diagrams lies in the competition between the tendency of electrons to delocalize and the pronounced interactions between charge, spin and lattice. We discuss here the nature of charge transport in the organic charge-transfer salt \alphaETI{}, a highly anisotropic material with a complex dielectric response underpinned by strong electron-electron interactions.

\begin{figure}
	\includegraphics[width=0.85\columnwidth,clip]{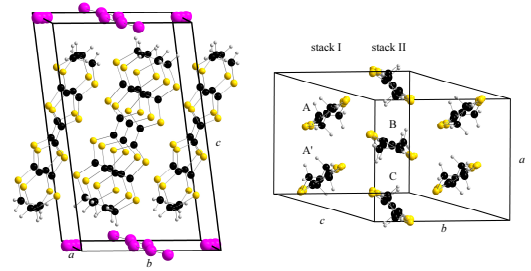}
	\caption{The crystal structure of \alphaETI{}. Carbon, sulphure, iodine and hydrogen atoms are represented by black, yellow, pink and grey spheres, respectively and the unit cell is marked with black lines. Crystallographic directions are marked as $a$, $b$ and $c$. Left: View along $a$ direction shows a typical layered structure where an organic BEDT-TTF layer is sandwiched between two inorganic I$_{3}$ layers. Right: View along $c$ direction shows a BEDT-TTF plane with the two-stack herring-bone pattern. Four different BEDT-TTF molecules in the unit cell are marked as A, A', B and C. This figure is based on data after Kakiuchi \etal{} \cite{Kakiuchi07}}
	\label{fig:structure}
\end{figure}

\alphaETI{}, where BEDT-TTF stands for bis\-(ethyleneditio)-tetrathiafulvalene, is the first organic material with highly conductive properties in two dimensions.\cite{Bender84-108} It is a layered structure of four BEDT-TTF molecules per unit cell organized in a planar two-stack herring-bone pattern (see Fig.\ \ref{fig:structure}). The molecules are separated by I$^-_3$ anions along the crystallographic $c^\ast$ direction. The resulting electronic properties are quasi-two-dimensional and strongly anisotropic. The phase diagram of \alphaETI{} features a number of intriguing quantum effects: at ambient pressure it undergoes a metal-to-insulator transition into a charge-ordered (CO) state at 135\,K with CO-induced ferroelectricity\cite{Takahashi06,Yamamoto08,Ivek12,Tomic15} where it shows ferroelectric hysteresis,\cite{Lunkenheimer15} nonlinear ultrafast optical response,\cite{Yamamoto08} and photo-induced phase transition,\cite{Iwai07,Peterseim16}
it features zero-gap semiconductivity with massless Dirac-like fermions,\cite{Tajima07,Tajima09,Peterseim16}
and becomes superconducting under uniaxial pressure.\cite{Tajima02}

At ambient pressure and high temperatures the system exhibits metallic character. Transport investigations showed that hydrostatic pressure reduces $T_\mathrm{CO}$.\cite{Kajita93,Ojiro93,Mishima95,Beyer16} Above 1.5\,GPa the CO transition is completely suppressed and the metallic region extends to low temperatures with an almost temperature-independent resistivity between 300 and 2\,K.\cite{Mishima95,Tajima06,Tajima00} Magnetotransport measurements above 1.5\,GPa indicate that in the same temperature range the carrier density and mobility change by about six orders of magnitude in such a manner that the effects just cancel out giving nearly constant resistivity. Below 4\,K, \alphaETI{} is in a state with a low carrier density of approximately $8 \times 10^{14} ~\mathrm{cm}^{-3}$ and an extremely high mobility of about $3\times 10^5\,\mathrm{cm}^2/\mathrm{Vs}$.\cite{Tajima06,Tajima00} Such unusual transport properties at high pressures are interpreted in terms of Dirac-cone type dispersion near the Fermi level\cite{Peterseim16} which was predicted by energy band calculations \cite{Kobayashi04,Katayama06} based on crystal structure under uniaxial strain. \cite{Kondo04} 

Density functional theory (DFT) calculations for \alphaETI{} under hydrostatic pressure \cite{Alemany12} predicted additional massive holes besides massless Dirac fermions, which has been confirmed by recent magnetotransport measurements under pressure.\cite{Monteverde13,Navarin15} According to the same calculations at ambient pressure, \cite{Alemany12} the high temperature phase of \alphaETI{} is semimetallic with small electron and hole pockets at the Fermi level. Even though the semimetallic nature of \alphaETI{} at ambient pressure was already hinted at more than thirty years ago by extended Hückel molecular orbital calculations,\cite{Mori84} direct experimental proof is still absent. Generally, a semimetallic state with electron and hole pockets is characterized by the existence of electrons and holes which have low density and high mobility. Hall effect and magnetoresistance measurements provide a powerful means to access the properties of both of these types of charge carriers. Concerning \alphaETI{} at ambient pressure, there is only one early magnetotransport publication but it did not take into account the two-carrier scenario, and it was limited to a simple quarter-filled metallic picture.\cite{Pokhodnya87}

Turning to the phenomenon of charge ordering in \alphaETI{} at ambient pressure, partial charge disproportionation and charge fluctuations are already present at room temperature and all the way down to the CO transition.\cite{Yue10,Moroto04} When temperature reaches below $T_\mathrm{CO}=135$\,K, a striped pattern of charge disproportionation sets in with approximately $+0.8e$, $+0.85e$, $+0.15e$ and $+0.2e$ charge per BEDT-TTF molecule.\cite{Takano01,Moldenhauer93,Dressel04,Kakiuchi07,Ivek11,Beyer16} Concomitantly, a gap opens in the spin and charge sector, making the system diamagnetic and insulating. Structural x-ray diffraction finds no superlattice reflections, meaning the lattice is not modulated and no Peierls-like electron-lattice coupling is responsible for the CO. However, a reduction of symmetry at $T_\mathrm{CO}$ from P$\bar{1}$ to P1 does take place.\cite{Kakiuchi07} It corresponds to a loss of inversion centers between molecules in stack I, the so-called molecules A and A$^\prime$, which allows for two domain types within the charge order.

The CO in \alphaETI{} is of a pronounced ferroelectric character.\cite{Kakiuchi07,Yamamoto08,Ivek10,Ivek11,Yamamoto10,Alemany12,Lunkenheimer15} According to structural analysis, the crystal shows monotonic lattice shrinkage without substantial displacement of molecules with lowering temperature, suggesting that structural modulation makes only a minor contribution to the electric polarization. Therefore the polarization was attributed mainly to the modulation of the electron distribution caused by CO.\cite{Yamamoto08} Below the CO transition an anisotropic dielectric relaxation in the radio-frequency range is observed.\cite{Ivek10,Ivek11,Lunkenheimer15} Within molecular planes, the dielectric spectra show a marked dispersion with two discernible contributions: the stronger one changes with temperature similarly to phason excitations in charge- and spin-density waves, whereas the smaller mode is temperature-independent and reminiscent of a soliton-like behavior.\cite{Ivek10,Ivek11} These features were used to describe the CO in \alphaETI{} as a cooperative bond-charge density wave with ferroelectric-like nature where both short-wavelength domain-wall excitations and long-wavelength phason-like excitations are present.\cite{Ivek11}. Even though the phenomenology is very similar to density wave systems, an open issue with such an interpretation is that it partially relies on a Peierls-like distortion of structure which is in particular absent in \alphaETI{}.\cite{Dressel94,Alemany12} A recent study of dielectric response in a temperature sweep perpendicular to molecular planes also found a similar dispersion, but arrived to a different interpretation.\cite{Lunkenheimer15} This picture stresses the short-range relaxor ferroelectricity which requires disorder. At this time the exact mechanism of the dielectric response both in-plane and out-of-plane is still not clarified.

The electronic behavior of \alphaETI{} is evidently still under heated discussion despite being subjected to decades of thorough research. In this paper we search for signatures and origin of disorder in the titular compound through a detailed systematic study of its dielectric properties, in-plane and out-of-plane, resistivity, Hall effect and magnetoresistance at ambient pressure. We present an interpretation of dielectric response that takes into account the intrinsic disorder in anion chains intimately coupled with BEDT-TTF molecules from stack I which are responsible for ferroelectricity. The same disorder causes a nearest-neighbor hopping contribution in dc transport and negative magnetoresistance. We further find that the high-temperature phase of \alphaETI{} is decidedly semimetallic and defined by a dominant inter-pocket scattering process, as well as evidence of strong fluctuations above the CO transition.

\section{Experimental}

Measurements were performed on flat, planar, high-quality single crystals of \alphaETI{} with typical dimensions of 3\,mm $\times$ 1\,mm $\times$ 0.5\,mm. The samples were oriented beforehand using mid-infrared reflectivity spectra. The in-plane $a$ and $b$ crystallographic axes correspond to the two directions of polarization which give extremal infrared reflectivity spectra. \cite{Ivek11,Ivek10,Dressel94,Pinteric14} The largest crystal surface is parallel to the molecular $ab$ planes, while the $c^\ast$-axis of the crystal corresponds to the direction perpendicular to the crystallographic $ab$ plane. Contacts for transport measurements were made by applying conductive carbon paint directly to the surface of the sample.

DC resistivity $\rho$ was measured by a standard four contact technique between room temperature and 25\,K along the three principal directions. The measurements were performed during both cooling and heating at rates 3--30\,K/h.  

Hall effect and magnetoresistance were measured in the temperature range $90\,\mathrm{K} < T < 300\,\mathrm{K}$ and in magnetic fields $B$ up to 9\,T. For all samples the current $I$ (5\,nA to 500\,$\mu$A) was applied along the $b$-axis and the magnetic field was oriented along the $c^\ast$-axis. Depending on the sample resistance, low-frequency ac (22\,Hz) or dc excitation was used. The measurements were performed at fixed temperatures in field sweeps from $-B_\mathrm{max}$ to $+B_\mathrm{max}$. In order to eliminate any possible influence of magnetoresistance, the Hall voltage was determined as $V_{xy} = [V_{xy}(+B)-V_{xy}(-B)]/2$. Hall coefficient $R_\mathrm{H}$ was then obtained as $R_\mathrm{H} = (V_{xy}t/IB)$, where $t$ is the sample thickness. The magnetoresistance was standardly determined as $\Delta\rho/\rho_{0} = [\rho(B)-\rho(0)]/\rho(0)$.

Temperature-dependent dielectric measurements were performed along the $a$- and $c^\ast$-axis from room temperature down to 4\,K and 20\,K, respectively, at frequencies in the kHz--MHz range using the Agilent 4294A impedance analyzer. At each temperature $T$ and frequency $\omega$ complex conductivity $\sigma(T, \omega)$ is measured and calculated to dielectric function $\varepsilon(T, \omega)$ using the standard expression $\varepsilon(T, \omega) = (\sigma(T, \omega)-\sigma(T, 0))/i\varepsilon_0\omega$ where $\varepsilon_0$ is the permittivity of vacuum. Special care was taken to subtract the background capacitance of the setup and sample holder, as well as to exclude any possible extrinsic effects due to sample preparation.\cite{Ivek10} Due to the metallic-like sample conductivity above $T_\mathrm{CO} = 135$\,K and a finite phase resolution of the impedance analyzer, reliable capacitance (imaginary conductivity) data was obtained only for temperatures in the insulating phase, \ie{}, below $T_\mathrm{CO}$.

\section{Results}


\begin{figure}
	\includegraphics[width=.85\columnwidth,clip]{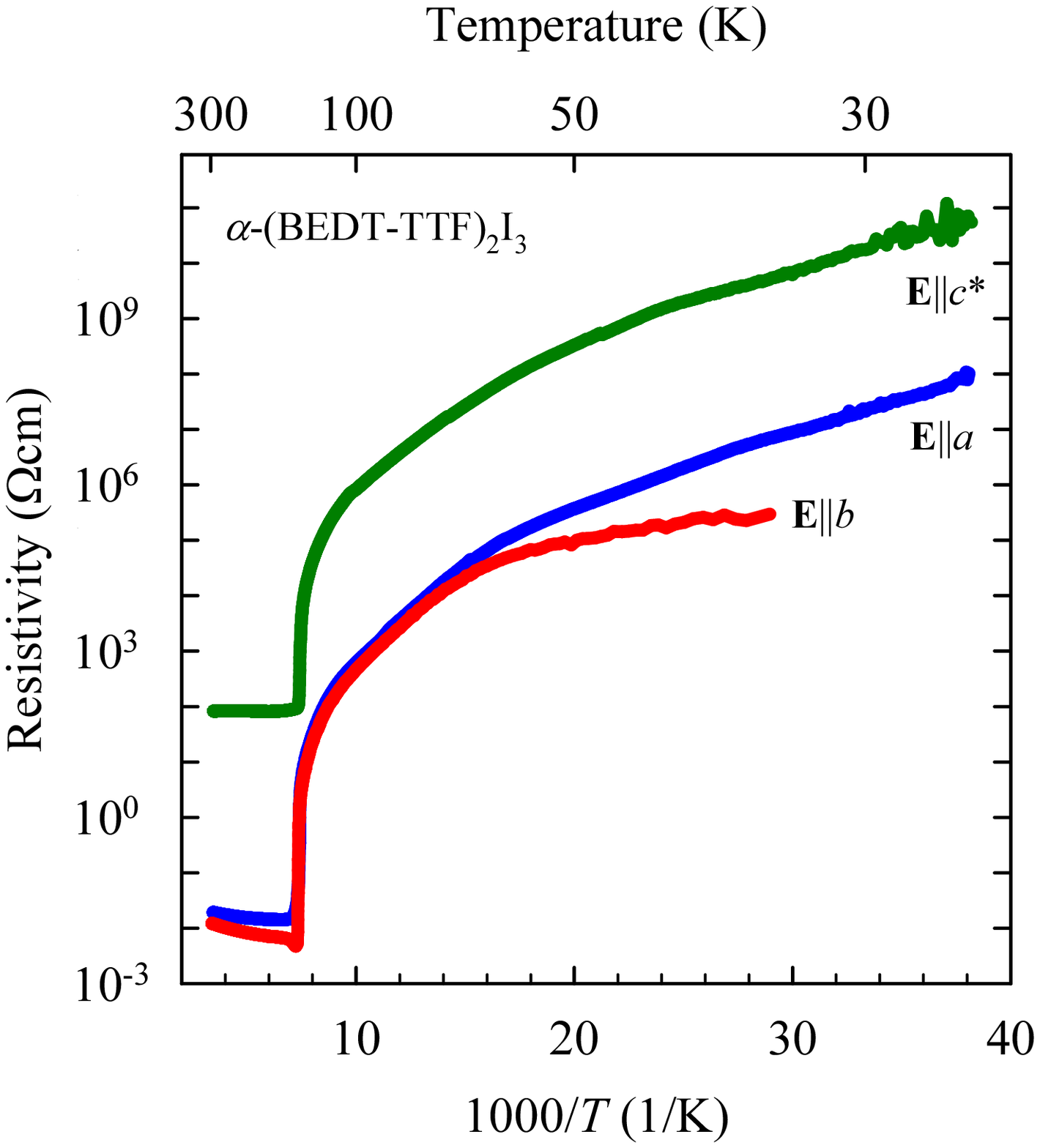}
	\caption{Temperature dependence of dc resistivities measured along the crystallographic $a$- (blue line), $b$- (red line) and $c^{\ast}$-axis (green line) for \alphaETI{}.}
	\label{fig:resistivity}
\end{figure}

Figure \ref{fig:resistivity} shows the temperature dependence of the resistivity of \alphaETI{} along the $a$-, $b$- and $c^\ast$-axis. Room temperature resistivities of the two in-plane directions $\rho_a = 19\,\mathrm{m\Omega{cm}}$ and $\rho_b = 12\,\mathrm{m \Omega {cm}}$ give the in-plane anisotropy $\rho_a/\rho_b = 1.6$, in good agreement with previously published results.\cite{Ivek11} The charge order transition is clearly visible as a sudden increase in the resistivity at $T_\mathrm{CO} = 135$\,K along all three directions. No histeretic behavior was found through thermal cycling between extremal temperatures. The resistivitiy along the $c^{\ast}$ direction, perpendicular to the BEDT-TTF planes, is three orders of magnitude larger than the in-plane resistivity in the whole temperature interval, in accord with the quasi-2D nature of the compound. The in-plane anisotropy remains nearly constant down to about 70\,K and then it starts to increase. Temperature dependence of the resistivity in the insulating phase cannot be described by one temperature-independent activation energy which opens the possibility of a temperature-dependent energy gap. Moreover, there is a sudden change of slope in the resistivity curve around 70\,K. This is most pronounced along $b$-direction, a strong indication of a complex transport mechanism in the charge-ordered state.

\begin{figure}
	\includegraphics[width=0.9\columnwidth,clip]{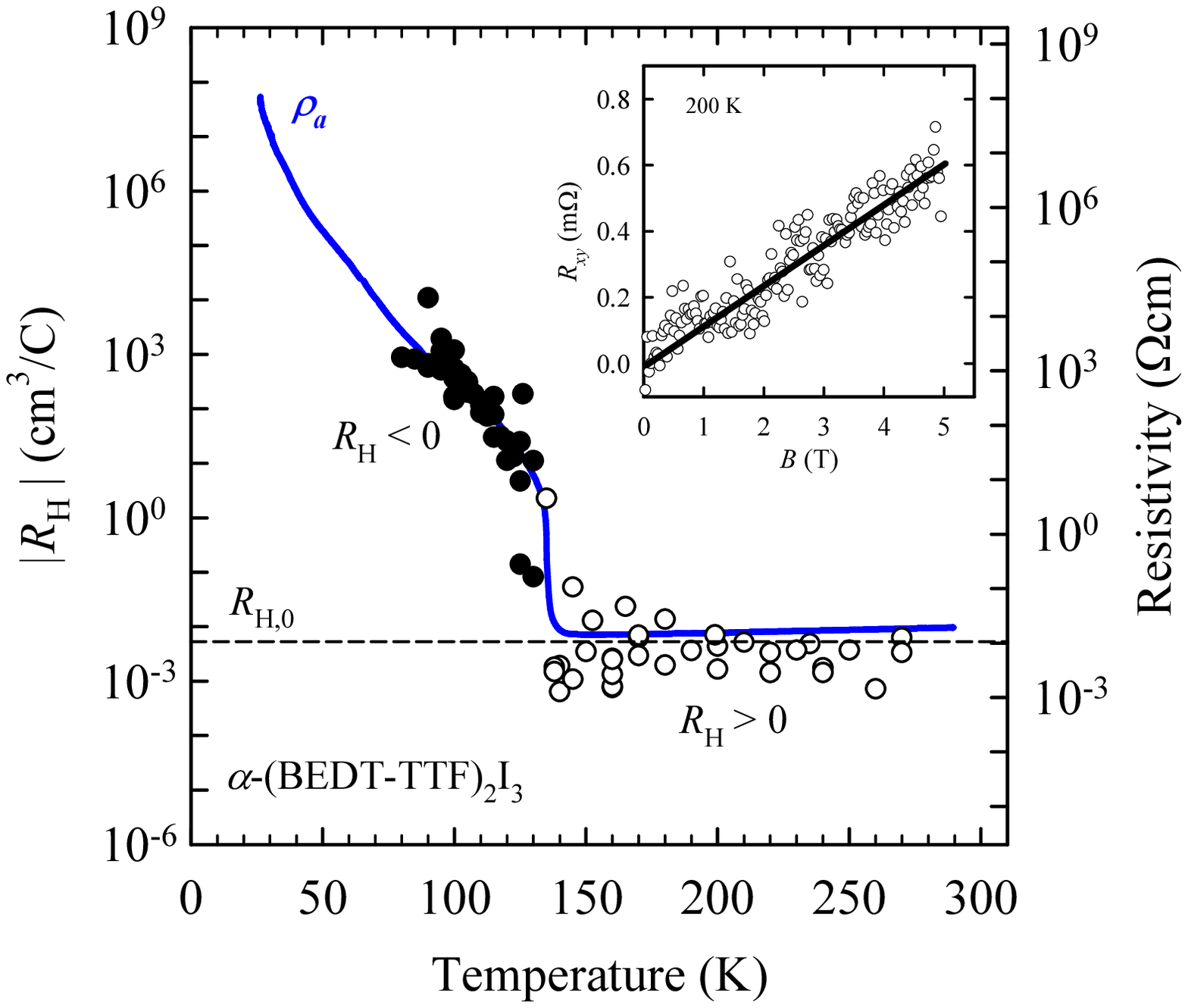}
	\caption{Temperature dependence of the Hall coefficient $R_\mathrm{H}$ (symbols) and resistivity $\rho_a$ (blue line) for \alphaETI{}. Empty and full symbols represent positive and negative values of $R_\mathrm{H}$, respectively. The black dashed line corresponds to the value calculated for a quarter-filled band (see text), $R_{\mathrm{H},0} = 5.29 \times 10^{-3}\,\mathrm{cm}^3/\mathrm{C}$. Inset: magnetic field dependence of the Hall resistance $R_{xy}$ is linear, a representative measurement is shown for $T = 200$\,K. Solid black line is the linear fit.} 
	\label{fig:Hall}
\end{figure}

Temperature dependence of the Hall coefficient $R_\mathrm{H}$ is shown by Fig.\ \ref{fig:Hall}. A simple charge transfer consideration between BEDT-TTF subsystem and iodine atoms leads to one hole per two BEDT-TTF molecules, \ie{}\ taking into account four molecules in the unit cell there are two holes per unit cell. Under the na\"{\i}ve assumption of four degenerate BEDT-TTF bands, these bands are quarter-filled by holes. This provides the rough estimate for the Hall coefficient $R_{\mathrm{H},0} = 1/e\,n_0 = +5.29 \times 10^{-3}$\,cm$^3/$C where $e$ is the electron charge and $n_0 = 2/V_\mathrm{cell} = 1.18\times 10^{21}$\,cm$^{-3}$ is calculated density of holes ($V_\mathrm{cell} = 1695.4$\,\AA{}$^3$). The measured Hall coefficient for $T > T_\mathrm{CO}$, $R_\mathrm{H} \approx 3 \times  10^{-3}$\,cm$^3$/C, is somewhat lower but not far from that value, and in agreement with previously published results (see Fig.\ \ref{fig:Hall}).\cite{Pokhodnya87} 

At $T_\mathrm{CO}$ the Hall coefficient abruptly changes sign and suddenly increases its absolute value which is a strong indication of a phase transition. Below $T_\mathrm{CO}$, $R_\mathrm{H}$ closely follows the temperature behavior of the dc resistivity. The Hall resistance $R_{xy} = V_{xy}/I$ is linear with magnetic field up to 9\,T in the whole temperature range. An example of its linearity is shown in the inset of Fig.\ \ref{fig:Hall}. 

\begin{figure}
\includegraphics[width=0.85\columnwidth,clip]{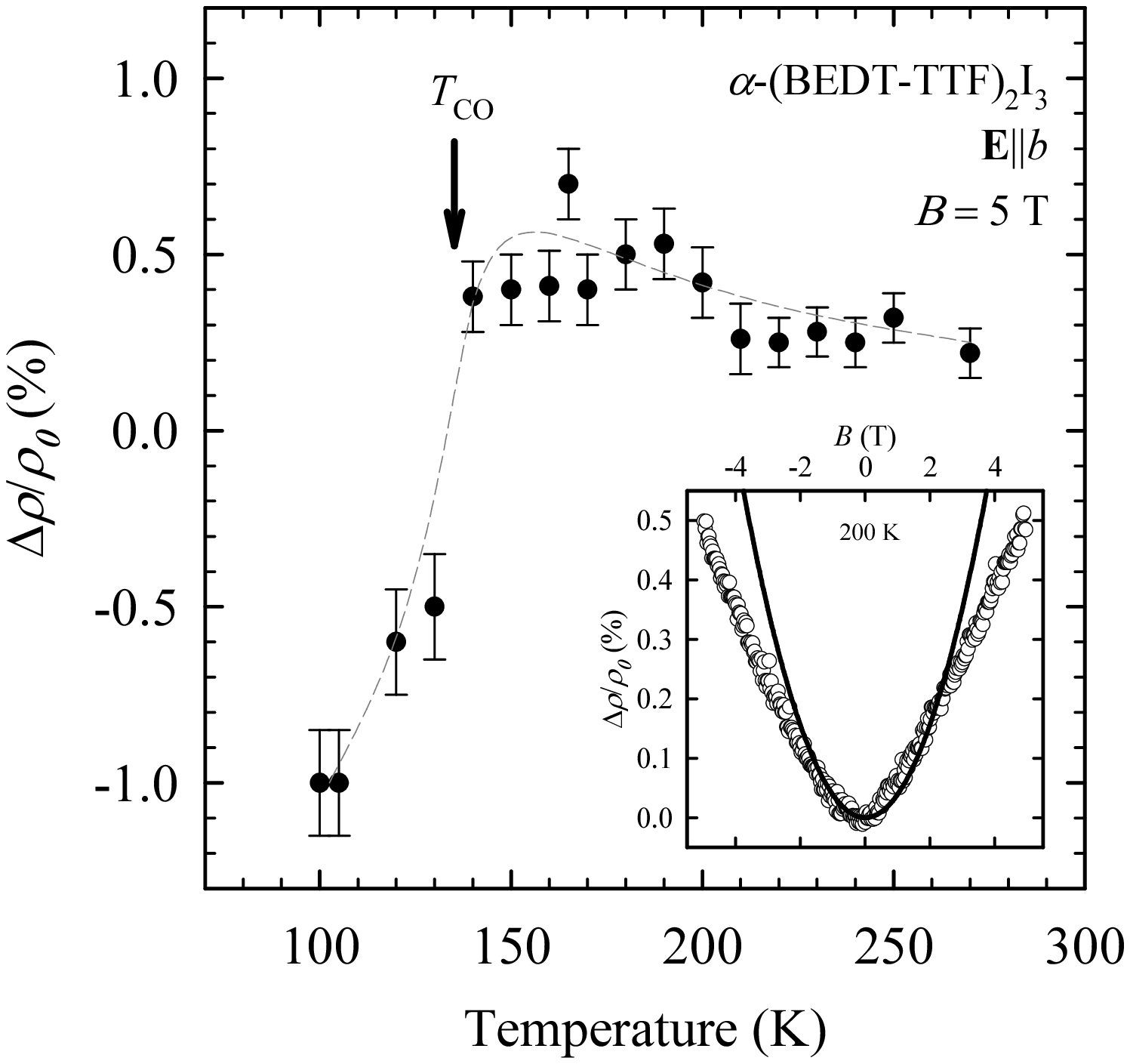}
\caption{The temperature dependence of the magnetoresistance $\Delta\rho/\rho_0$ calculated for $B = 5$\,T for \alphaETI{}. Dashed line is a guide for the eye. Inset: magnetic field dependence of $\Delta\rho/\rho_{0}$ at $T = 200$\,K. Full line is the $B^2$ dependence which is valid up to 2\,T.} 
\label{fig:magnetoresistance}
\end{figure}

Figure \ref{fig:magnetoresistance} shows the temperature dependence of the magnetoresistance $\Delta\rho/\rho_0$ at $B = 5$\,T. $\Delta\rho/\rho_0$ at low fields follows a $B^2$ dependence, while in the $B \geq 2$\,T regime it increases more slowly (see inset of Fig.\ \ref{fig:magnetoresistance}). Magnetoresistance is positive above $T_\mathrm{CO}$ with the average value near the room temperature around 0.3\,\%. It slightly increases with lowering temperature and finally below $T_\mathrm{CO}$ changes sign and becomes negative.

\begin{figure}
\includegraphics[width=0.8\columnwidth,clip]{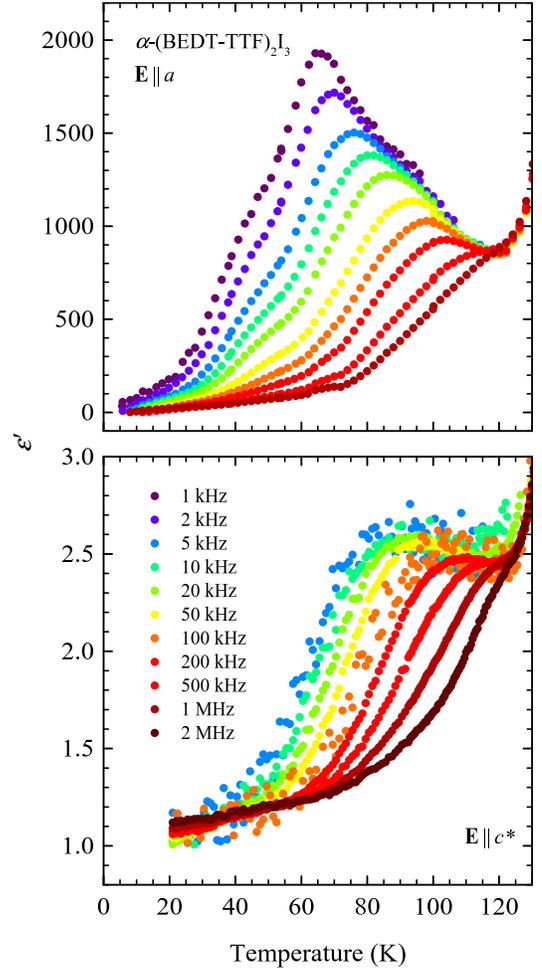}
\caption{Temperature dependence of real part of dielectric response $\varepsilon^\prime$ for various frequencies, in-plane (upper panel) and out-of-plane (lower panel).} 
\label{fig:ds_T_sweep}
\end{figure}

Figure \ref{fig:ds_T_sweep} presents the strongly anisotropic real part of dielectric function $\varepsilon^\prime$ as a function of temperature. Well below $T_\mathrm{CO}$, the in-plane direction $\mathbf{E}||a$ (upper panel) features a pronounced peak at all measured frequencies. The high-temperature shoulders of peaked curves align to follow the same curve, the behavior which is characteristic for relaxor ferroelectrics. In fact, for the in-plane direction a second, smaller peak can be resolved in the temperature sweep of $\varepsilon^\prime$. This particular detailed structure seen in \alphaETI{} is due to the two-mode response in frequency space; the two modes were previously assigned to the long-wavelength so-called phason-like mode, and soliton-like relaxation of the charge order.\cite{Ivek10,Ivek11}

The out-of-plane dielectric response is almost three orders of magnitude weaker in strength which correlates with the anisotropy of dc conductivity. No clear Curie-like peak is visible at $T_\mathrm{CO}$ and a relaxor-like peak starts to form below 70\,K, its appearance somewhat limited by a restricted frequency range. The out-of-plane results are in good agreement with previous work.\cite{Lunkenheimer15}

\section{Discussion and analysis}

\subsection{Charge carrier mobilities and the importance of inter-pocket scattering}

In the high-temperature phase, the experimentally determined Hall coefficient $R_\mathrm{H}$ for $T > T_\mathrm{CO}$ apparently supports the quarter-filled band picture. However, if we calculate the mobility of holes near room temperature in this single carrier metallic picture, $\mu = \sigma R_\mathrm{H} \approx 0.5 ~\mathrm{cm}^2/\mathrm{Vs}$, we get the upper bound on magnetoresistance\cite{Beer_book} $(\mu B)^2 \approx10^{-7}$ which is five orders of magnitude smaller than the measured values at $B = 5$\,T. A single-carrier metallic picture cannot describe the high-temperature phase of \alphaETI{}. As an additional argument, it is well known that the magnetoresistance of organic conductors at high temperatures in metallic state is usually undetectable.\cite{KorinHamzic03} DFT \cite{Alemany12} calculations for $T > T_\mathrm{CO}$ predict a semimetallic state with small electron and hole pockets at the Fermi level, so a two-carrier model presents itself as a natural fit. Conductivity, Hall coefficient and magnetoresistance of a two-carrier system are generally given by convoluted expressions which contain four unknown quantities, mobilities of electrons $\mu_\mathrm{e}$, holes $\mu_\mathrm{h}$, as well as their densities $n_\mathrm{e}$ and $n_\mathrm{h}$. Those expressions can be significantly simplified if we assume the stoichiometric compound, i.e.\ if the material may be considered sufficiently clean so the self-doping is not pronounced, with equal densities of electrons and holes $n_\mathrm{e} = n_\mathrm{h} = n$:\cite{Fawcett_two_carrier}
\begin{equation}
\label{eq:conductivity}
\sigma = e n(\mu_\mathrm{e} + \mu_\mathrm{h}),
\end{equation}
\begin{equation}
\label{eq:RH}
R_\mathrm{H} = \frac{1}{e n} \frac{\mu_\mathrm{h} - \mu_\mathrm{e}}{\mu_\mathrm{e} + \mu_\mathrm{h}},
\end{equation} 
\begin{equation}
\label{eq:MR}
\Delta\rho/\rho_{0} = \mu_\mathrm{e} \mu_\mathrm{h} B^2.
\end{equation} 
Although expressions (\ref{eq:conductivity}), (\ref{eq:RH}) and (\ref{eq:MR}) are derived for isotropic bands, they can be used to describe a quasi-2D system. Namely, in our measurements the magnetic field is parallel to $c^{\ast}$-axis and as such forces charge carriers to move in the $ab$ plane where transport at high temperatures is almost isotropic (see Figure \ref{fig:resistivity}). Thus, from the conductivity, Hall effect, and magnetoresistance data one can in principle determine the mobilities and densities of the charge carriers and separate conductivity contributions coming from electrons and holes. Such an analysis previously explained unusual transport properties of \alphaETI{} at high pressures \cite{Tajima06} and recently a colossal magnetoresistance in a perfectly compensated semimetal WTe$_{2}$. \cite{Ali2014} 

In nice agreement with Equation (\ref{eq:RH}), $R_\mathrm{H}$ does not depend on the magnetic field $B$, \ie{}\ Hall resistance is linear with respect to magnetic field (see inset of Figure \ref{fig:Hall}). At the same time, Equation (\ref{eq:MR}) gives $\Delta\rho/\rho_0 \propto B^{2}$ which holds up to 2\,T, a sizable range if we consider the elevated temperatures. Taking the magnetoresistance data up to 2\,T we can then obtain a rough estimate of charge carrier mobilities and densities. 

Above $T_\mathrm{CO}$ we obtain charge carrier densities $n_\mathrm{e} = n_\mathrm{h} \approx 10^{18}\,\mathrm{cm}^{-3} \approx 0.002/$cell, a value which does not depend on temperature and is three orders of magnitude smaller than calculated for a quarter-filled band (2/cell) as in the above text. Such a low density of carriers is in accord with distinctly small electron and hole pockets at the Fermi level instead of a metal-like, large Fermi surface. The carrier mobilities turn out to be almost the same and around 200\,cm$^2$/Vs which is at least two orders of magnitude larger than typical values in organic conductors\cite{Tajima06,KorinHamzic90,Saeki12,Culo15} and much closer to values found in other semimetals.\cite{Lovett_book} Note that within the limitations of our model the $R_\mathrm{H}>0$ is due to $\mu_\mathrm{h} - \mu_\mathrm{e}>0$ where the difference in mobilities is significantly below 1\%. Thus, the analysis of magnetotransport data shows that above $T_{\mathrm{CO}}$ \alphaETI{} is a system with high mobility and low density of charge carriers and gives a strong experimental confirmation of the semimetallic phase.

It is a puzzling contradiction that small electron and hole pockets result in a Hall coefficient close to the one expected from a quarter-filled band. According to Equation (\ref{eq:RH}), the electron and hole contributions partially cancel each other out and therefore the effective $R_\mathrm{H}$ can end up much smaller than expected from carrier density alone. 

The same densities and mobilities of electrons and holes would require a complete compensation in $R_\mathrm{H}$. The situation is very similar to 1T-TiSe$_{2}$, \cite{Velebit16} another layered material with charge order where the compensation of electron and hole contributions greatly reduces $R_\mathrm{H}$ above the charge-ordering transition. The very close values of mobilities for electrons and holes in both compounds indicate a very specific scattering process. 

Previous work on the electronic structure of \alphaETI{} in the high-temperature phase agrees on two bands barely crossing the Fermi level at different points in the Brillouin zone. \cite{Alemany12} The high-temperature phase with one band almost full and the other almost empty (valence and conduction band, respectively) puts \alphaETI{} squarely among the indirect-gap semimetals. The corresponding Fermi surface consists of small electron and hole pockets. The respective concentrations of charge carriers are equal in the stoichiometric compound and much smaller than what is obtained within a simple picture of quarter-filled bands.

The mobilities obtained from our measurements are large compared to other organic conductors, so let us attempt a microscopic explanation. Energy and momentum conservation allows only scattering processes that move carriers inside the corresponding pocket (intra-pocket scattering) and the ones that transfer carriers between the two pockets (inter-pocket scattering). Considering the size of pockets, this restriction strongly reduces electron-phonon scattering which results in large relaxation times and consequently high mobilities of the charge carriers. The relative strength of intra- and inter-pocket scattering depends on the electron-phonon couplings for all the phonon branches involved. Without detailed calculations it is hard to argue which of the two scattering processes, if any, is dominant. However, the inter-pocket scattering should be particularly favorable in indirect-gap semimetals, as the energies involved are low (quasi-elastic on the electrons scale), whereas the momentum transfer is always large. In this way impurities as well as phonon bands both contribute to scattering processes that are efficient in reducing electric current.

The inter-pocket scattering has a particular property which conspires to equalize the mobilities of carriers in the two pockets. Namely, the carriers in the hole pocket are scattered into the electron pocket, and vice versa. In both cases the scattering rates are determined by the density of final states. Therefore, the scattering rate of holes is proportional to the electron band mass $\gamma_\mathrm{h} \propto g^2 m_\mathrm{e}$, and vice versa the scattering rate of electrons is proportional to the hole band mass $\gamma_\mathrm{e} \propto g^2 m_\mathrm{h}$ ($g$ denotes the relevant electron-phonon coupling for the inter-pocket scattering). This leads to the hole mobility being same as the electron mobility, 
\begin{equation}
	\mu_\mathrm{h} \approx C \frac{1/\gamma_\mathrm{h}}{m_\mathrm{h}} = C^\prime \frac{1}{m_\mathrm{h} g^2 m_\mathrm{e}} = C \frac{1/\gamma_\mathrm{e}}{m_\mathrm{e}} \approx \mu_\mathrm{e},
\end{equation}
assuming other scattering channels being negligible (the proportionality factors $C$ and $C^\prime$ contain some less interesting factors and are introduced for convenience).

This scenario is consistent with our analysis of magnetotransport properties in \alphaETI{} for $T > T_\mathrm{CO}$ which showed that the electron and hole mobilities not only have high values but also coincide to a great precision. The vanishing difference in the mobilities of electron and holes suggests that intra-pocket scattering is indeed very small in comparison to the contribution of the inter-pocket scattering.   

As the carrier densities remain constant, the change of conductivity above $T_\mathrm{CO}$ is accounted for solely by the change of mobilities. However, in the low-temperature phase a complete analysis cannot be performed because a negative magnetoresistance is not taken into account by the Equation (\ref{eq:MR}). At the CO phase transition $R_\mathrm{H}$ changes sign (see Fig.\ \ref{fig:magnetoresistance}). A negative magnetoresistance is rare in non-magnetic materials and usually has an exotic origin: in organics it has been ascribed to band splitting which induces a small increase of charge carrier density,\cite{Tiedje75} 2D weak localization due to disorder in the anion lattice,\cite{Ulmet88} or to reduced scattering on antiferromagnetic fluctuations.\cite{Basletic96} Negative magnetoresistance is also observed in the impurity conduction of many semiconductors.\cite{Fritzsche55,Woods64,Sasaki66} Thus, apart from being just one of the indications of the CO phase transition we can take the negative magnetoresistance in \alphaETI{} as a signal for presence of disorder. Indeed, we shall see in the following that the influence of disorder is also present in dc transport in the form of hopping contribution, as well as in the relaxor-like dielectric response.

\subsection{Evolution of the low-temperature transport gap in presence of disorder}

\begin{figure}
	\includegraphics[width=.85\columnwidth,clip]{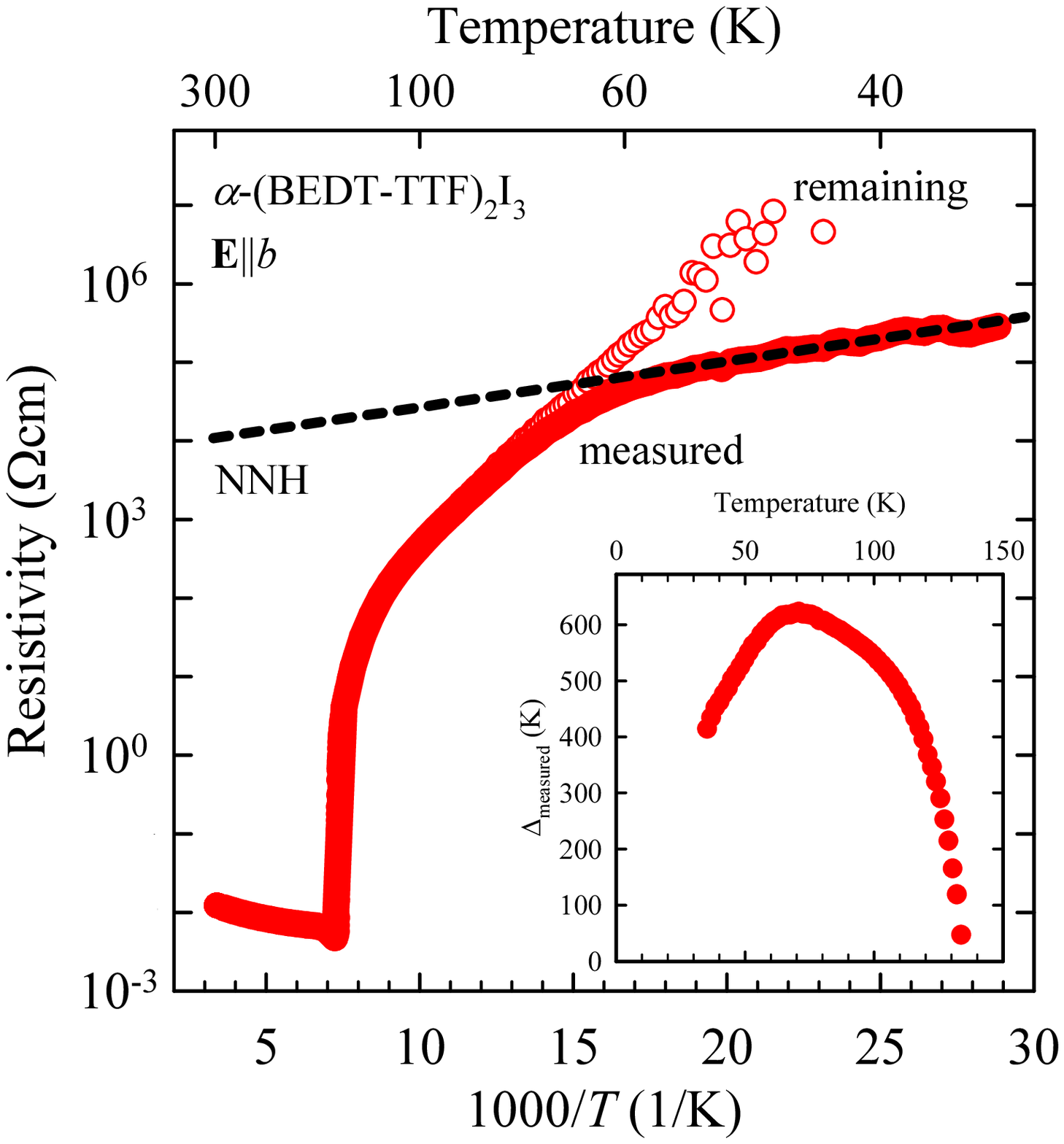}
	\caption{Decomposition of the measured resistivity of \alphaETI{} along $b$ direction (full red circles) into two conductivity channels: the nearest-neighbor hopping channel (black dashed line) and the mean-field-like channel (empty red circles). Inset: temperature-dependent activation energy calculated directly from measured resistivity using the Equation (\ref{eq:delta_vs_T}) for $\mathbf{E}||b$. The apparent activation energy increases from $T_\mathrm{CO}$ down to 70\,K, but then starts to decrease towards lower temperatures, which indicates two transport channels are present (see text).}
	\label{fig:resistivity_correction}
\end{figure}

The resistivity curves in the insulating charge-ordered state all show a temperature-dependent slope in the Arrhenius plot (Fig.\ \ref{fig:resistivity}). Evidently, a simple activation law
\begin{equation}
\rho(T) = \rho_0 \exp(\Delta/T)
\label{eq:simple_activation}
\end{equation}
with a temperature-independent activation energy $\Delta$ is not appropriate. Further, in potentially disordered systems one option is to consider a variable range hopping mechanism.\cite{Mott_book,Efros_VRH} However, following the procedure outlined by Joung \etal{},\cite{Joung12} we find it does not describe well any of the three measured directions. Let us instead consider an activated transport mechanism with a general, temperature-dependent activation energy $\Delta(T)$ which is then simple to extract from experimental data: 
\begin{equation}
\label{eq:delta_vs_T}
\Delta(T) = T \cdot \ln{\left(\rho(T)/\rho_0\right)}.
\end{equation}
If the constant $\rho_0$ is set in such a way that the activation energy $\Delta$ vanishes at $T_\mathrm{CO}$, we obtain the temperature dependence of activation energy shown by Figure \ref{fig:resistivity_correction} (inset) for $\mathbf{E}||b$. $\Delta$ starts to increase at $T_\mathrm{CO}$ down to about 70\,K and then it decreases.

According to the activation energy analysis in TMTTF family of compounds and its relation to the energy gap, \cite{Rose_TMTTF} a decrease of activation energy at low temperatures may be caused by some form of charge carrier hopping between disorder-induced localized states at the Fermi level. Seeing that variable-range hopping is ruled out, we ascribe the low temperature conductivity mechanism in \alphaETI{} to a nearest-neighbor hopping (NNH). The NNH channel requires randomly distributed, localized states of a certain density which implies a level of disorder present in all our measured samples.

\begin{figure}
	\includegraphics[width=0.9\columnwidth]{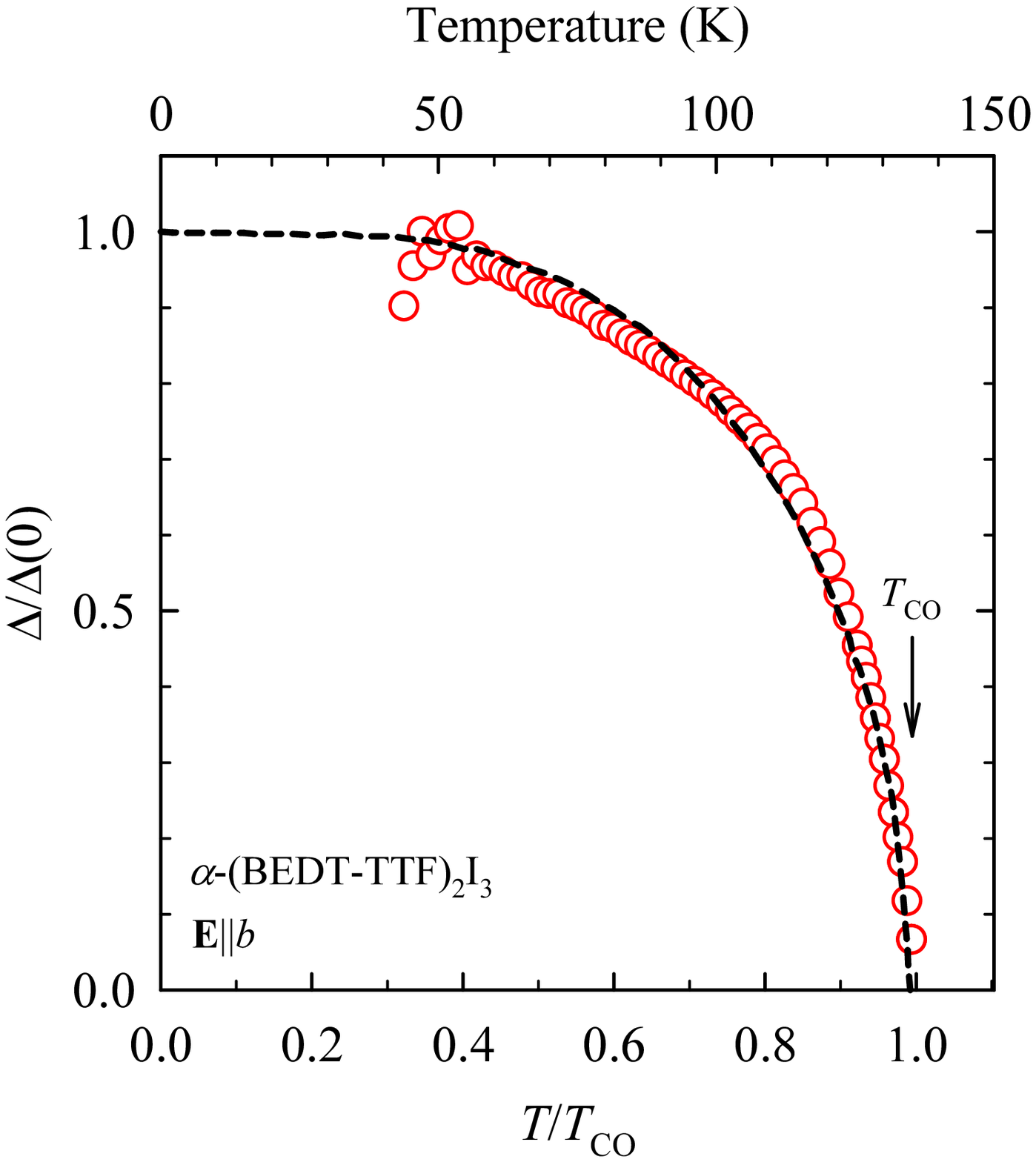}
	\caption{Temperature dependence of the normalized transport activation energy $\Delta$ in \alphaETI{} for a representative direction $\mathbf{E}||b$ (empty red circles). $\mathbf{E}||a$ and $\mathbf{E}||c^{\ast}$ show the same temperature behavior and are omitted for clarity. Dashed line represents the normalized temperature dependence of the mean-field theoretical order parameter.\cite{Tinkham_book}} 
	\label{fig:delta_vs_T}
\end{figure}

The NNH is most often modeled as a simple activated behavior with a temperature-independent activation energy as in Equation (\ref{eq:simple_activation}).\cite{Gantmakher_book} Assuming there is a second charge transport mechanism present parallel to the NNH, our measured resistivity curves for all three directions can be decomposed as
\begin{equation}
\label{eq:two_channels}
1/\rho_\mathrm{measured} = 1/\rho_\mathrm{NNH} + 1/\rho_\mathrm{remaining}.
\end{equation}
Figure \ref{fig:resistivity_correction} shows the decomposition of resistivity to two contributions for the representative direction $\mathbf{E}||b$. Inserting the remaining non-NNH contribution $\rho_\mathrm{remaining}(T)$ into Equation (\ref{eq:delta_vs_T}) we get the temperature dependence of its activation energy shown in Figure \ref{fig:delta_vs_T}. It appears to be strikingly mean-field-like down to the lowest temperatures along all three crystallographic directions. The activation energy at 0\,K, $\Delta(0)$, can be determined by fitting the theoretical mean-field curve to our data:\cite{Tinkham_book} it is isotropic and a value around 700\,K is obtained for all three measured directions. Hence, we can associate our $2\Delta(T)$ with a mean-field-like energy gap which evolves continuously below the 3D CO phase transition.

A comment is in order on the relation between the transport gap and the measured optical gap.\cite{Ivek11,Beyer16} The former develops with temperature and converges to $2\Delta(0) \approx 1400$\,K, while the latter opens abruptly at $T_\mathrm{CO}$ and was reported to be $2\Delta_\mathrm{CO}\approx870$\,K. One would expect transport to happen across the smallest available gap at a given temperature. Having the optical, direct gap smaller than the transport gap points to a possible issue in determining these gaps. Close to the optical gap there are phonon features (around 800\,\cm{} and 1300\,\cm{}) that impede its reliable determination, so its actual value may lie somewhat higher. Further, the effective dc transport gap (double the activation energy in the inset of Fig.\ \ref{fig:resistivity_correction} extracted directly from measured resistivity) never goes above about 1200\,K. This brings the apparent disagreement between the transport and the optical gap closer to resolution. According to DFT calculations\cite{Alemany12} the smallest energy difference between the valence and conduction band needs to be indirect. Therefore, we can associate our transport gap with an indirect transition and the optical gap with a direct transition near the Dirac-like point. 

The transport energy gap $2\Delta(T)$ follows a mean-field behavior, but the ratio $2\Delta(0)/T_\mathrm{CO} \approx 10$ is far from 3.5 expected from the conventional BCS mean-field theory. Similarly high values of $2\Delta(0)/T_{c}$ were observed in low-dimensional systems and ascribed to one of the following: suppression of the mean-field transition temperature $T_\mathrm{c}$ by strong coupling interactions,\cite{Monceau_book,Rossnagel11} imperfect nesting,\cite{Huang_PRB42} or fluctuations pronounced in low-dimensional systems.\cite{Monceau_book} Here the latter seems most relevant as fluctuating charge is observed in \alphaETI{} already at room temperature.\cite{Yue10} According to Kup\v{c}i\'{c} \etal{}\cite{Kupcic14} fluctuations have little influence on the low-temperature value of the theoretical order parameter $2\Delta(0)$, but substantially lower the critical temperature and thus strongly affect the $2\Delta(0)/T_\mathrm{c}$ ratio. DC transport in the CO phase of \alphaETI{} therefore points towards a predominantly mean-field-like behavior. 

The charge ordering at $T_\mathrm{CO}$ carries certain signatures of a first-order phase transition as seen by specific heat,\cite{Fortune} charge disproportionation,\cite{Takano01,Kakiuchi07,Ivek11} and opening of the optical gap.\cite{Ivek11} On the other hand, in the CO phase some experimental quantities evolve in a way which suggests a second-order transition, such as the BCS-like behavior of magnetic susceptibility,\cite{Rothaemel86} intensity of second harmonic generation,\cite{Yamamoto08} as well as the dc transport activation energy reported here. This apparent conflict might be understood if the abrupt first-order transition sets the stage for a continuous change of these quantities below $T_\mathrm{CO}$. Further work is certainly needed in order to fully explain the complex nature of the CO phase transition in \alphaETI{}.

\subsection{The origin of disorder in \alphaETI{}}
The nearest-neighbor hopping in dc data and the negative magnetoresistance implicate the presence of disorder in BEDT-TTF conducting layers. Some related compounds feature disordered ethylene groups of BEDT-TTF molecules that can significantly influence the low-temperature state.\cite{Pinteric02} But, in \alphaETI{} the ethylene groups seem to be ordered at all temperatures.\cite{REF_no_ethy_disorder_in_alpha} To identify the source of disorder we turn to anionic layers. Indeed, we remind that already at room temperature early x-ray diffraction measurements noticed intense diffuse lines which were assigned to disorder within the I$^-_3$ anion chains.\cite{Dressel94} There are hydrogen bonds between anions and the ethylene groups of BEDT-TTF molecules on A, A$^\prime$, and B sites which influence the concentration of holes in the highest-occupied molecular orbitals,\cite{Alemany12} so it is plausible to regard the I$^-_3$ displacement as intrinsic disorder which directly influences charge transport. Moreover we can associate this disorder with the unusual properties found by infrared electronic conductivity,\cite{Ivek11} namely the in-plane non-zero optical conductivity within the gap area and its gradual evolution with cooling even to the lowest temperatures.

In the end let us address the origin of the dielectric relaxation in the CO phase of \alphaETI{}. From optical second harmonic generation it is known that the metal-to-insulator phase transition is evidently of ferroelectric character, long-range and three-dimensional. In a long-range ferroelectric system a frequency-independent Curie-like peak is expected, such as in the well-established CO-driven ferroelectric (TMTTF)$_2$AsF$_6$.\cite{Nad06} However, in \alphaETI{} only at lower temperatures, below about 100\,K, does a wide, dispersive $\varepsilon^\prime$ appear both in-plane and out-of-plane. In disordered systems, a relaxor-like response would commonly be taken as evidence of glassy physics with freezing, short-range, polar entities.\cite{Cross87} One signature of such a response is the broad distribution of mean relaxation times, commonly described by the value $1-\alpha$, which approaches 0.3 close to the freezing temperature. This type of process is observed in the related $\kappa$-(BEDT-TTF)$_2X$ systems where no long-range charge ordering is found.\cite{Pinteric14,Pinteric16} For \alphaETI{}, an interpretation of this kind\cite{Lunkenheimer15} would be in contradiction with the long-range order.

We offer here a revised picture for the dielectric response which reconciles disparate interpretations. The ferroelectricity of \alphaETI{} comes from the charge-rich and charge-poor molecules A and A$^\prime$ with broken inversion symmetry.\cite{Kakiuchi07,Alemany12} Coincidentally, these two molecular sites also have the most hydrogen bonds with I$^-_3$ anions.\cite{Alemany12} It is therefore plausible that the I$^-_3$ disorder\cite{Dressel94} will strongly manifest itself within the response of the bulk polarization, meaning the large peak in $\varepsilon'$ directly stems from the bulk. With this in mind we propose that the apparent relaxor-like behavior observed both in-plane and out-of-plane is effectively a ``renormalized'' Curie response associated with the CO transition and the bulk polarization: in the presence of disorder it is being pushed to temperatures well below $T_\mathrm{CO}$ and becomes dispersive. Now, the dispersion is determined by free carrier screening, since the mean dielectric relaxation time follows the same temperature dependence as dc resistivity.\cite{Ivek10,Ivek11} Further, the distribution width of dielectric relaxation times extracted from frequency domain experiments is $1-\alpha\approx 0.7$--0.8,\cite{Ivek10,Ivek11} too narrow for a disordered, glassy system with freezing relaxor response. In the presence of disorder, dielectric response with such a width and screening is a well-known fingerprint of incommensurate density waves.\cite{Gruener88} We posit that in \alphaETI{} the intrinsic anion disorder plays a role analogous to screened pinning on the incommensurate potential in density waves. Lastly, since this interpretation leaves the long-range ferroelectric ordering in place, we associate the remaining smaller dielectric contribution with domain wall motion.

\section{Conclusion}
In this work we address the nature of charge transport and charge ordering in \alphaETI{}. In the high-temperature semimetallic phase, transport is governed by large mobilities of electrons and holes. The value of the Hall coefficient for this semimetal unexpectedly corresponds to that of a quarter-filled band, and we explain this surprising feature as a consequence of a predominantly inter-pocket scattering which equalizes mobilities of the two types of charge carriers. At low temperatures the dc transport points toward two separate channels of conduction: a nearest-neighbor hopping contribution and an activated contribution with a mean-field, distinctly isotropic behavior. In dielectric response, the ferroelectric nature of the charge-ordered phase is announced by an anisotropic, dispersion stemming from the bulk response, without any sign of freezing of dielectric moments. Our measurements indicate that the intrinsic anion disorder influences ac and dc transport in the molecular conducting layers through hydrogen bonds with BEDT-TTF molecules. The issue of order of the phase transition and its relation to the mean-field-like transport behavior remains to be clarified.

\begin{acknowledgments}
We thank J.\ P.\ Pouget and I.\ Kup\v{c}i\'{c} for enlightening discussions. The work has been supported by the Croatian Science Foundation project IP-2013-11-1011 and by the Deutsche Forschungsgemeinschaft (DFG).
\end{acknowledgments}


\begin{thebibliography}{99}
	
\bibitem{Bender84-108}
K.\ Bender, I.\ Hennig, D.\ Schweitzer, K.\ Dietz, H.\ Endres, and H.\ J.\ Keller,
Mol.\ Cryst.\ Liq.\ Cryst.\ {\bf 108}, 359 (1984).

\bibitem{Kakiuchi07}
T.\ Kakiuchi, Y.\ Wakabayashi, H.\ Sawa, T.\ Takahashi, and T.\ Nakamura,
J.\ Phys.\ Soc.\ Jpn.\ {\bf 76}, 113702 (2007).

\bibitem{Takahashi06}
T.\ Takahashi, Y.\ Nogami, and K.\ Yakushi,
J.\ Phys.\ Soc.\ Jpn.\ {\bf 75}, 051008 (2006).

\bibitem{Yamamoto08}
K.\ Yamamoto, S.\ Iwai, S.\ Boyko, A.\ Kashiwazaki, F.\ Hiramatsu, C.\ Okabe, N.\ Nishi, and K.\ Yakushi,
J.\ Phys.\ Soc.\ Jpn.\ {\bf 77}, 074709 (2008).

\bibitem{Ivek12}
T.\ Ivek, I.\ Kova\v{c}evi\'{c},\ M.\ Pinteri\'{c},  B.\ Korin-Hamzi\'{c}, S.\ Tomi\'{c}, T.\ Knoblauch, D.\ Schweitzer, and M.\ Dressel,
Phys.\ Rev.\ B {\bf 86}, 245125 (2012).

\bibitem{Tomic15}
S.\ Tomi\'{c} and M.\ Dressel, Rep.\ Prog.\ Phys.\ {\bf 78}, 096501 (2015).

\bibitem{Lunkenheimer15}
P.\ Lunkenheimer, and A.\ Loidl, J.\ Phys:\ Condens.\ Matter\ {\bf 27}, 373001 (2015).

\bibitem{Iwai07}
S.\ Iwai, K.\ Yamamoto, A.\ Kashiwazaki, F.\ Hiramatsu, H.\ Nakaya, Y.\ Kawakami, K.\ Yakushi, H.\ Okamoto, H.\ Mori, and Y.\ Nishio,
Phys.\ Rev.\ Let..\ {\bf 98}, 097402 (2007).

\bibitem{Peterseim16}
T.\ Peterseim, T.\ Ivek, D.\ Schweitzer, and M.\ Dressel,
Phys.\ Rev.\ B {\bf 93} 245133 (2016).

\bibitem{Tajima07}
N.\ Tajima, S.\ Sugawara, M.\ Tamura, R.\ Kato, Y.\ Nishio, and K.\ Kajita,
Eur.\ Phys.\ Lett.\ {\bf 80}, 47002 (2007).

\bibitem{Tajima09}
N.\ Tajima, S.\ Sugawara, R.\ Kato, Y.\ Nishio, and K.\ Kajita,
Phys.\ Rev.\ Lett.\ {\bf 102}, 176403 (2009).

\bibitem{Tajima02}
N.\ Tajima, A.\ Ebina-Tajima, M.\ Tamura, Y.\ Nishio, and K.\ Kajita,
J.\ Phys.\ Soc.\ Jpn. {\bf 71}, 1832 (2002).

\bibitem{Kajita93}
K.\ Kajita, T.\ Ojiro, H.\ Fujii, Y.\ Nishio, H.\ Kobayashi, A.\ Kobayashi, and R. Kato, J.\ Phys.\ Soc.\ Japan\ {\bf 61}, 23 (1993).

\bibitem{Ojiro93}
T.\ Ojiro, K.\ Kajita, Y.\ Nishio, H.\ Kobayashi, A.\ Kobayashi, R.\ Kato, and Y.\ Iye, Synth.\ Met.\ {\bf 56}, 2268 (1993).

\bibitem{Mishima95}
T.\ Mishima, T.\ Ojiro, K.\ Kajita, Y.\ Nishio, and Y.\ Iye, Synth.\ Met.\ {\bf 70}, 771 (1995).

\bibitem{Beyer16}
R.\ Beyer, A.\ Dengl, T.\ Peterseim, S.\ Wackerow, T.\ Ivek, A.\ V.\ Pronin, D.\ Schweitzer, and M.\ Dressel, Phys.\ Rev.\ B {\bf 93}, 195116 (2016).

\bibitem{Tajima06}
N.\ Tajima, S.\ Sugawara, M.\ Tamura, Y.\ Nishio, and K.\ Kajita, J.\ Phys.\ Soc.\ Japan\ {\bf 75}, 051010 (2006).

\bibitem{Tajima00}
N.\ Tajima, M.\ Tamura, Y.\ Nishio, K.\ Kajita, and Y.\ Iye, J.\ Phys.\ Soc.\ Japan\ {\bf 69}, 543 (2000).

\bibitem{Kobayashi04}
A.\ Kobayashi, S.\ Katayama, K.\ Noguchi, and Y.\ Suzumura, J.\ Phys.\ Soc.\ Japan\ {\bf 73}, 3135 
(2004).

\bibitem{Katayama06}
S.\ Katayama, A.\ Kobayashi, and Y.\ Suzumura, J.\ Phys.\ Soc.\ Japan\ {\bf 75}, 054705 (2006).

\bibitem{Kondo04}
R.\ Kondo, and S.\ Kagoshima, J.\ Phys.\ IV\ France\ {\bf 114}, 523 (2004).

\bibitem{Alemany12}
P.\ Alemany, J.-P.\ Pouget, and E.\ Canadell, Phys.\ Rev.\ B\ {\bf 85}, 195118 (2012).

\bibitem{Monteverde13}
M.\ Monteverde, M.\ O.\ Goerbig, P.\ Auban-Senzier, F.\ Navarin, H.\ Henck, C.\ R.\ Pasquier, C.\ M\'ezi\` ere, and P.\ Batail, Phys.\ Rev.\ B\ {\bf 87}, 245110 (2013).

\bibitem{Navarin15}
F.\ Navarin, E.\ Tisserond, P.\ Auban-Senzier, C.\ M\'ezi\` ere, P.\ Batail, C.\ Pasquier, and M.\  Monteverde, Physica\ B\ {\bf 460}, 257 (2015).

\bibitem{Mori84}
T.\ Mori, A.\ Kobayashi, Y.\ Sasaki, H.\ Kobayashi, G.\ Saito, and H.\ Inokuchi, Chem.\ Lett.\ {\bf  1984}, 957 (1984).

\bibitem{Pokhodnya87}
K.\ I.\ Pokhodnya, Y.\ V.\ Sushko, and M.\ A.\ Tanatar, Sov.\ Phys.\ JETP\ {\bf 65}, 795 (1987).

\bibitem{Yue10}
Y.\ Yue, K.\ Yamamoto, M.\ Uruichi, C.\ Nakano, K.\ Yakushi, S.\ Yamada, T.\ Hiejima, and A.\ Kawamoto,
Phys.\ Rev.\ B {\bf 82}, 075134 (2010).

\bibitem{Moroto04}
S.\ Moroto, K.-I.\ Hiraki, Y.\ Takano, Y.\ Kubo, T.\ Takahashi, H.\ M.\ Yamamoto, and T.\ Nakamura, J.\ Phys.\ IV France {\bf 114}, 339 (2004).

\bibitem{Takano01}
Y.\ Takano, K.\ Hiraki, H.\ M.\ Yamamoto, T.\ Nakamura, and T.\ Takahashi,
J.\ Phys.\ Chem. Solids {\bf 62}, 393 (2001).

\bibitem{Moldenhauer93}
J.\ Moldenhauer, C.\ Horn, K.\ I.\ Pokhodnia, D.\ Schweitzer, I.\ Heinen, and H.\ J.\ Keller,
Synth.\ Met.\ {\bf 60}, 31 (1993).

\bibitem{Dressel04}
M.\ Dressel and N.\ Drichko,
Chem.\ Rev.\ {\bf 104}, 5689 (2004).

\bibitem{Ivek11}
T.\ Ivek, B.\ Korin-Hamzi\'{c}, O.\ Milat, S.\ Tomi\'{c}, C.\ Clauss, N.\ Drichko, D.\ Schweitzer, and M.\ Dressel,
Phys.\ Rev.\ B {\bf 83}, 165128 (2011).

\bibitem{Ivek10}
T.\ Ivek, B.\ Korin-Hamzi\'{c}, O.\ Milat, S.\ Tomi\'{c}, C.\ Clauss, N.\ Drichko, D.\ Schweitzer, and M.\ Dressel, Phys.\ Rev.\ Lett.\ {\bf 104}, 206406 (2010).

\bibitem{Yamamoto10}
K.\ Yamamoto, A.\ A.\ Kowalska, and K.\ Yakushi, Appl.\ Phys.\ Lett.\ {\bf 96}, 122901 (2010).

\bibitem{Dressel94}
M.\ Dressel, G.\ Gr\"{u}ner, J.\ P.\ Pouget, A.\ Breining, and D.\ Schweitzer,
J.\ de Physique I (France) {\bf 4}, 579 (1994);
Synth.\ Met.\ {\bf 70}, 929 (1995).

\bibitem{Pinteric14}
M.\ Pinteri\'{c}, M.\ \v{C}ulo, O.\ Milat, M.\ Basleti\'{c}, B.\ Korin-Hamzi\'{c}, E.\ Tafra, A.\ Hamzi\'{c}, T.\ Ivek, T.\ Peterseim, K.\ Miyagawa, K.\ Kanoda, J.\ A.\ Schlueter, M.\ Dressel, and S. Tomi\'{c},
Phys.\ Rev.\ B {\bf 90}, 195139 (2014).

\bibitem{Beer_book}
A.\ C.\ Beer, {\it Solid State Physics suppl. 4, ed. F. Seitz and D. Turnbull} (Academic Press, New York, 1963).

\bibitem{KorinHamzic03}
B.\ Korin-Hamzi\'{c}, E.\ Tafra, M.\ Basleti\'{c}, A. Hamzi\'{c}, G.\ Untereiner, and M.\ Dressel, 
Phys.\ Rev.\ B 67, {\bf 014513} (2003).

\bibitem{Fawcett_two_carrier}
E.\ Fawcett, Adv.\ in\ Physics\ {\bf 13}, 139 (1964).

\bibitem{Ali2014}
M.\ N.\ Ali, J.\ Xiong, S.\ Flynn, J.\ Tao, Q.\ D.\ Gibson, L.\ M.\ Shoop, T.\ Liang, N.\ Haldolaarachchige, M.\ Hirschberger, N.\ P.\ Ong, and R.\ J.\ Cava, Nature {\bf 514}, 205 (2014).

\bibitem{KorinHamzic90}
B.\ Korin-Hamzi\'{c}, L.\ Forr\'{o}, and J.\ R.\ Cooper, Phys.\ Rev.\ B {\bf 41}, 11646 (1990).

\bibitem{Saeki12}
A.\ Saeki, Y.\ Koizumi, T.\ Aida, and S.\ Seki, Acc.\ Chem.\ Res.\ {\bf 45},  1193 (2012).

\bibitem{Culo15}
M.\ \v{C}ulo, E.\ Tafra, M.\ Basleti\'{c}, S.\ Tomi\'{c}, A.\ Hamzi\'{c}, B.\ Korin-Hamzi\'{c}, M.\ Dressel, and J.\ A.\ Schlueter, Physica B {\bf 460}, 208 (2015).

\bibitem{Lovett_book}
D.\ R.\ Lovett, {\it Semimetals and Narrow-Bandgap Semiconductors.} (Pion Limited, 207 Brondesbury Park, London NW2 5JN, 1977).

\bibitem{Velebit16}
K.\ Velebit, P.\ Pop\v{c}evi\'{c}, I.\ Batisti\'{c}, M.\ Eichler, H.\ Berger, L.\ Forr\'{o}, M.\ Dressel, N.\ Bari\v{s}i\'{c}, and E.\ Tuti\v{s},Phys.\ Rev.\ B {\bf 94}, 075105 (2016).

\bibitem{Tiedje75}
T.\ Tiedje, J.\ F.\ Carolan, and A.\ J.\ Berlinsky, Can.\ J.\ Phys.\ {\bf  53}, 1593 (1975).

\bibitem{Ulmet88}
J.\ P.\ Ulmet, L.\ Bachere, S.\ Askenazy, and J.\ C.\ Ousset, Phys.\ Rev.\ B {\bf 38}, 7782 (1988).

\bibitem{Basletic96}
M.\ Basleti\'c, D.\ Zanchi, B.\ Korin-Hamzi\'c, A.\ Hamzi\'c, S.\ Tomi\'c, J.\ M.\ Fabre, J.\ Phys.\ I\ France\ {\bf 6}, 1855 (1996). 

\bibitem{Fritzsche55}
H.\ Fritzsche, and K.\ Lark-Horovitz, Phys.\ Rev.\ {\bf 99}, 400 (1955).

\bibitem{Woods64}
J.\ F.\ Woods, and  C.\ Y.\ Chen, Phys.\ Rev.\ {\bf 135}, A1462 (1964).

\bibitem{Sasaki66}
W.\ Sasaki, J.\ Phys.\ Soc.\ Japan\ {\bf 21} Suppl., 543(1966).

\bibitem{Mott_book}
N.\ F.\ Mott\, and E.\ A.\ Davis, {\it Electronic Processes in Non-crystalline Solids} (Oxford University Press, London, 1971).

\bibitem{Efros_VRH}
A.\ L.\ Efros, and B.\ I.\ Shklovskii, J.\ Phys.\ C\ {\bf 8}, L49 (1975).

\bibitem{Joung12}
D.\ Joung and S.\ I.\ Khondaker, Phys.\ Rev.\ B {\bf 86}, 235423 (2012).

\bibitem{Rose_TMTTF}
B.\ K\"{o}hler, E.\ Rose, M.\ Dumm, G.\ Untereiner, and M. Dressel, Phys.\ Rev.\ B\ {\bf 84}, 035124 (2011).

\bibitem{Gantmakher_book}
V.\ F.\ Gantmakher, {\it Electrons and Disorder in Solids} (OUP, Oxford, 2005).

\bibitem{Tinkham_book}
M.\ Tinkham, {\it Introduction to Superconductivity} (McGraw-Hill Kogakusha, Ltd., Tokyo, 1975), p.\ 35.

\bibitem{Monceau_book}
P.\ Monceau (Ed.), {\it Electronic Properties of Inorganic Quasi-One-Dimensional Compounds} (D. Reidel Publishing Co., Dordrecht, Holland, 1985).

\bibitem{Rossnagel11}
K.\ Rossnagel, J.\ Phys.: Condens.\ Matter {\bf 23}, 213001 (2011).

\bibitem{Huang_PRB42}
X.\ Huang, and K.\ Maki, Phys.\ Rev.\ B\ {\bf 42}, 6498 (1990).

\bibitem{Kupcic14}
I.\ Kup\v{c}i\'{c}, Z.\ Rukelj, and S.\ Bari\v{s}i\'{c}, J.\ Phys.:\ Condens.\ Matter\ {\bf 26}, 195601 (2014).

\bibitem{Fortune}
N.\ A.\ Fortune,  K.\ Murata, M.\ Ishibashi, M.\ Tokumoto, N.\ Kinoshita, and H.\ Anzai,  Solid State Commun. {\bf 77}, 265 (1991).

\bibitem{Rothaemel86}
B.\ Rothaemel, L.\ Forr\'{o}, J.\ R.\ Cooper, J.\ S.\ Schilling, M.\ Weger, P.\ Bele, H.\ Brunner, D.\ Schweitzer, and H.\ J.\ Keller,
Phys.\ Rev.\ B {\bf 34}, 704 (1986).

\bibitem{Pinteric02}
M.\ Pinteri\'{c}, S.\ Tomi\'{c}, M.\ Prester, \DJ{}.\ Drobac, and K.\ Maki,
Phys.\ Rev.\ B {\bf 66}, 174521 (2002).

\bibitem{REF_no_ethy_disorder_in_alpha}
E.\ Canadell, private communication (2014).

\bibitem{Nad06}
F.\ Nad and P.\ Monceau, Jour.\ Phys.\ Soc.\ Jpn. {\bf 75}, 051005 (2006).

\bibitem{Cross87}
L.\ E.\ Cross, Ferroelectrics {\bf 76}, 241 (1987).

\bibitem{Pinteric16}
M.\ Pinteri\'{c}, P.\ Lazi\'{c}, A.\ Pustogow, T.\ Ivek, M. Kuve\v{z}di\'{c}, O.\ Milat, B.\ Gumhalter, M.\ Basleti\'{c}, M.\ \v{C}ulo, B.\ Korin-Hamzi\'{c}, A.\ L\"{o}hle, R.\ H\"{u}bner, M.\ Sanz Alonso, T.\ Hiramatsu, Y.\ Yoshida, G.\ Saito, M.\ Dressel, and S. Tomi\'{c},
Phys.\ Rev.\ B {\bf 94}, 161105(R) (2016).

\bibitem{Gruener88}
G.\ Gr\"{u}ner, Rev.\ Mod.\ Phys.\ {\bf 60}, 1129 (1988); {\bf 66}, 1 (1994).








\end{thebibliography}
\end{document}